\documentclass[twocolumn,showpacs,preprintnumbers,amsmath,amssymb,prb]{revtex4}

\usepackage{graphicx}
\usepackage{dcolumn}
\usepackage{bm}
\usepackage{amsmath}

\begin{document}


\title{Local gating of a graphene Hall bar by graphene side gates}

\author{F. Molitor}
\email{fmolitor@phys.ethz.ch}
\author{J. G\"uttinger}
\author{C. Stampfer}
\author{D. Graf}
\author{T. Ihn}
\author{K. Ensslin}
\affiliation{Solid State Physics Laboratory - ETH Zurich, Switzerland}

\date{\today}

\begin{abstract}
We have investigated the magnetotransport properties of a single-layer graphene Hall bar with additional graphene side gates. The  side gating in the absence of a magnetic field can be modeled by considering two parallel conducting channels within the Hall bar. This results in an average penetration depth of the side gate created field of approx. 90~nm. The side gates are also effective in the quantum Hall regime, and allow to modify the longitudinal and Hall resistances.
\end{abstract}

\maketitle



\section{Introduction}
The discovery of graphene flakes consisting of a single atomic layer \cite{novoselov_pnas2005} facilitated measurements of the so-called half-integer quantum Hall effect. \cite{novoselov_nature2005, zhang_nature2005} The electron and hole densities in planar structures were tuned using a back gate electrode. In order to achieve a lateral modulation of the charge density, additional laterally patterned gate electrodes are required. This has been achieved by covering the graphene flake with an isolating dielectric, and then depositing metallic top gates on the dielectric.\cite{lemme_condmat2007,williams_science2007,huard_prl2007} The procedure includes additional processing steps, the dielectric material needs to be of very high quality in order to avoid charge traps, and for very small structures the alignment of the structure and top gates could be difficult. In order to fabricate nanostructures on graphene, such as quantum wires, ribbons, quantum point contacts, and quantum dots more sophisticated gating schemes are required. This is also one of the requirements for the realization of spin qubits in graphene. \cite{trauzettel_naturephys2007} Narrow graphene ribbons have been realized by etching techniques. \cite{han_prl2007,chen_condmat2007} Lateral metallic gates have been employed \cite{graf_prb2007} to tune the electron distribution in a mesoscopic few-layer graphite wire.

Side gates have become extremely useful for tunable electronic nanostructures realized for example in AlGaAs heterostructures. \cite{luscher_apl1999} Here we set out to explore the possibility to use graphene side gates as a tuning knob for the electron density in a neighboring narrow graphene Hall bar. Side gates and Hall bar are processed from the same graphene flake facilitating a single-step technological process. A voltage applied to all side gates shifts the resistance curve measured versus back gate voltage, and increases the minimum conductivity value. These effects can be described with a simple model which allows to extract a characteristic penetration depth of the lateral field. We also investigate the case of a p-n configuration along the Hall bar. The data can be described using the same model and parameters. The charge distribution tuned by side gate voltages is also investigated in the quantum Hall regime.


\section{Sample and setup}

Mechanical exfoliation of natural graphite flakes and subsequent deposition onto substrates produces flakes with a great variety of thicknesses, sizes and shapes. \cite{novoselov_pnas2005} Among them are flakes with a height of one single atomic layer. They can be unambiguously distinguished from bilayer and thicker flakes by Raman spectroscopy. \cite{gupta_nanolett2006,ferrari_prl2006,graf_nanolett2007} The single layer flake presented in this paper was produced with this technique. It lies on a highly doped silicon substrate covered by 295~nm of thermal oxide. The oxide layer has two purposes: it facilitates the detection of flakes of single layer thickness by optical microscopy \cite{blake_apl2007}, and it allows the use of the highly doped silicon as a global back gate. The geometry was defined by electron beam lithography using 90~nm of PMMA as resist, followed by a reactive ion etching step based on argon and oxygen (9:1). Cr/Au (5~nm/60~nm) ohmic contacts were added in a second step.

\begin{figure}
\includegraphics[width=0.45\textwidth]{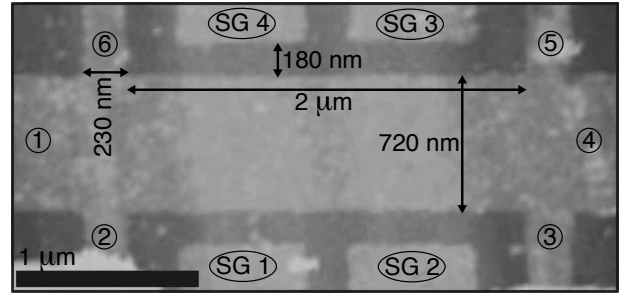}
\caption{\label{fig:Hallbar} Scanning force micrograph of the Hall bar with graphene side gates. Numbers 1-6 indicate contact pads for transport measurement. Electrodes SG 1 to SG 4 label the side gates.}
\end{figure}

The scanning force micrograph in Fig. \ref{fig:Hallbar} shows the etched structure, a single layer Hall bar with four additional graphene side gates. The Hall bar has a width of $\approx720$~nm, a length between the voltage probes of $\approx2~\mu$m, and the gap to the side gates is $\approx180$~nm. The mobility is $\mu \approx$5000~cm$^{2}$V$^{-1}$s$^{-1}$ at $T$=2~K.

Measurements were done at $T$=2~K in a four-terminal setup by applying an AC current through the outer contacts, indicated by numbers 1 and 4 in Fig.~\ref{fig:Hallbar}. The longitudinal resistance is determined by measuring the voltage difference between contacts 2 and 3, and the Hall resistance is determined from the voltage difference measured between contacts 3 and 5. 


\section{Results and discussion}


\subsection{Electric field effect}\label{sec:efe}

\begin{figure}
\includegraphics[width=0.45\textwidth]{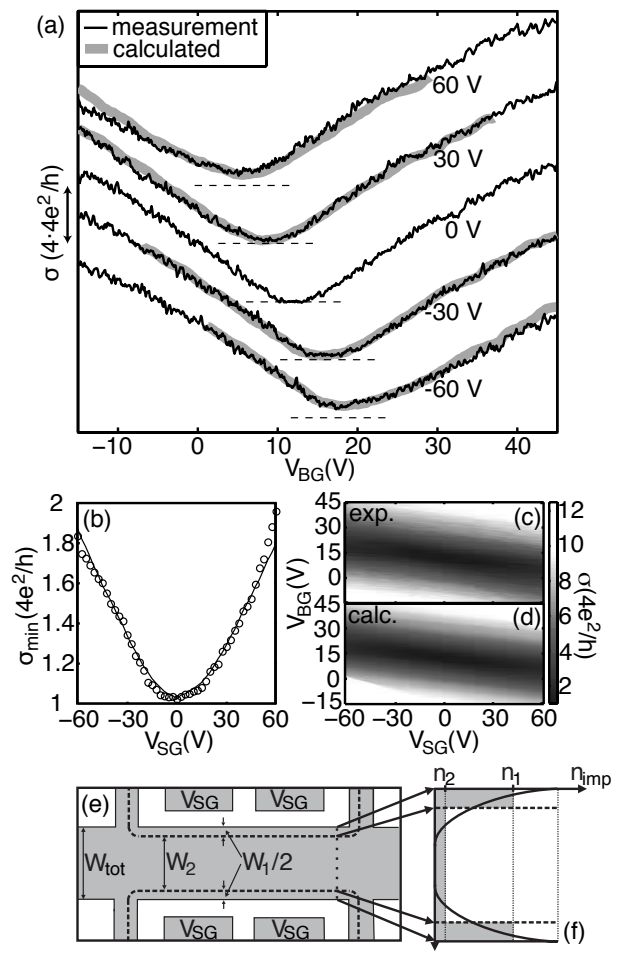}
\caption{\label{fig:allSG_0T} (a) Conductivity versus back gate voltage $V_{\mathrm{BG}}$ for various settings of the side gate voltage $V_{\mathrm{SG}}$ as indicated.  The same voltage is applied to all four side gates, and the measurement is performed with an ac current of 20~nA rms. The black curves represent the measured traces, while the grey lines are calculated with the model explained in section~\ref{sec:SG} taking into account two parallel conducting channels with different influence from the side gates. (b) Value of the minimum in conductivity as a function of applied side gate voltage, determined from the smoothed measured curves (circles), and calculated traces (line). (c),(d) Greyscale plot of the conductivity as a function of applied side gate and back gate voltage, (c) representing the measured date, (d) the calculations. (e) True to scale schematic of the Hall bar, with illustration of the two channels considered in the model. (f) Schematic section of the induced carrier density over the width of the Hall bar, with indication of the two regions (grey areas), the simplified  step-like density distribution used in the model and a realistic distribution (solid line).}
\end{figure}

In a standard metallic system, the field effect is irrelevant due to the extremely short screening length. However, by creating atomically thin graphene films, it is possible to tune the carrier density with a global back gate, and even to switch between electron and hole transport. \cite{novoselov_science2004} Conductivity traces as a function of back gate voltage are displayed in Fig.~\ref{fig:allSG_0T}(a) (black curves). The different traces correspond to a varying voltage applied to all four side gates simultaneously. The curves are vertically offset for clarity by 4$\cdot$4e$^{2}$/h. The dashed lines indicate a conductivity value of 4e$^{2}$/h. At both sides of the minimum in conductivity, where electron and hole densities compensate each other, the conductivity increases almost linearly with applied back gate voltage, since $n=\alpha(V_{\mathrm{BG}}-V_{\mathrm{BG,CN}})$, where $\alpha=7.2\cdot10^{10}$~cm$^{2}$/V and $V_{\mathrm{BG,CN}}$ is the position of the charge neutrality point in back gate voltage. \cite{novoselov_nature2005}

Applying a voltage $V_{\mathrm{SG}}$ to all four side gates has mainly two effects: The position of the conductivity minimum in back gate voltage shifts according to the relative lever arm. The minimum in conductivity, which has a value $\sigma_{\mathrm{min}}\approx 4e^{2}/h$ for $V_{\mathrm{SG}}$=0~V, increases with both positive and negative side gate voltages. A grayscale plot of the conductivity as a function of back and side-gate voltages  is displayed  in Fig. \ref{fig:allSG_0T}(c). Dark colors correspond to low conductivity.


\subsection{Model for the side gate influence}\label{sec:SG}
The back gate resembles a quasi-infinite plane with respect to the extent of the Hall bar. For large back gate voltages the carrier density in the Hall bar could thus become inhomogeneous because of the lateral and vertical electric field distribution. We neglect this effect since the distance between the Hall bar and the back gate electrode is smaller than the width of the Hall bar $W_{\mathrm{tot}}$.
The confining potential for the electrons perpendicular to the current direction of the Hall bar is expected to be rather steep on the scale of the Hall bar width. The density profile with an applied side gate voltage looks schematically as the solid line in Fig. \ref{fig:allSG_0T}(f). In the Hall bar center the density is homogeneous and given by $n_{0}$, whereas closer to the Hall bar edges, the density profile increases due to the action of the side gates. In order to simplify the analysis we approximate the induced density by a step function which gives rise to  two parallel conducting channels: two thin regions of width $W_{1}/2$ and density $n_{1}$ along the two edges of the Hall bar, where the side gates have a strong influence, and a broad channel of width $W_{2}=W_{\mathrm{tot}}-W_{1}$ and density $n_{2}$ in the center of the Hall bar, being only slightly influenced by an applied side gate voltage [Fig. \ref{fig:allSG_0T}(e)]. Inside each of these regions, a side gate voltage is assumed to induce a homogeneous carrier density, proportional to the applied voltage [Fig. \ref{fig:allSG_0T}(f)].

First we assume that for $V_{\mathrm{SG}}=0$~V [middle trace in Fig. \ref{fig:allSG_0T}(a)] all regions have the same density. 
We proceed by regarding the channel widths ($W_{1}, W_{2}$), as well as their lever arms ($\alpha _{1}, \alpha _{2}$)  with respect to the side gates as parameters. Together with the constraint $W_{\mathrm{tot}}=W_{1}+W_{2}=720~\mathrm{nm}$ this results in three free parameters.

We further assume that the conductivity inside each of these channels as a function of $V_{\mathrm{BG}}$ for a fixed $V_{\mathrm{SG}}$ equals the trace recorded for $V_{\mathrm{SG}}=0$~V, shifted by a back gate voltage value proportional to the potential shift induced by the side gates in this region. This gives the following expression for the conductivity: 
\begin{multline}
\sigma(V_{\mathrm{SG}},V_{\mathrm{BG}})=\beta \cdot \sigma(0,V_{\mathrm{BG}}-\alpha _{1}V_{\mathrm{SG}})\\+(1-\beta) \cdot \sigma(0,V_{\mathrm{BG}}-\alpha _{2}V_{\mathrm{SG}}),
\end{multline}
where $\beta=W_{1}/W_{\mathrm{tot}}$ is the relative width of region 1, and $\alpha _{1}$ and $\alpha _{2}$ are the relative average lever arms of the side gates in region 1 and 2. 

The parameters $\alpha_{1}$, $\alpha_{2}$ and $\beta$ are determined by searching the best agreement for the curve measured at $V_{\mathrm{SG}}=-60~\mathrm{V}$. The resulting parameters are  4/15, 1/12, 1/4. This corresponds to an average penetration depth of the electric field created by the side gates into the Hall bar of $W_{1}/2\approx 90$~nm. This width characterizes the area where the lever arm of the side gate is significant. These parameter values are then used for the calculation of the conductivity for all the other side gate voltages without further free parameters.

The gray lines in Fig. \ref{fig:allSG_0T}(a) represent the results of this calculation for different side gate voltages, along with the measured curves (black lines). The model reproduces the conductivity traces in the full range of applied side and back gate voltages. A complete picture for all the different side gate voltage values is shown in the greyscale plot of Fig. \ref{fig:allSG_0T}(d), representing the calculated conductivity as a function of $V_{\mathrm{SG}}$ and $V_{\mathrm{BG}}$, and can be compared to the corresponding measured data [Fig. \ref{fig:allSG_0T}(c)]. The white corners result from the limited back gate voltage range recorded for $V_{\mathrm{SG}}=0~\mathrm{V}$. 

The minimum conductivity value in graphene at the charge neutrality point has been widely discussed. While most theories predict an universal minimum, in some cases at the commonly observed value of $4e^{2}/h$, \cite{novoselov_nature2005} recent self-consistent calculations predict the value of the minimum to depend on the concentration of charged impurities. \cite{adam_condmat2007} 
Fig. \ref{fig:allSG_0T}(b) analyzes the minimum conductivity value as a function of the applied side gate voltage. The circles represent the minima of the measured traces, and the line the values determined from the model described above. Without any applied side gate voltage, the minimum conductivity lies at the commonly expected value of $4e^{2}/h$. An applied side gate voltage increases this value, until it almost reaches $8e^{2}/h$ in the case of $V_{\mathrm{SG}}=\pm60~\mathrm{V}$. This shows that it is possible to increase the minimum value of conductivity by creating a density gradient along the width of the conductor. 


\subsection{Side gates in p-n configuration}

\begin{figure}
\includegraphics[width=0.45\textwidth]{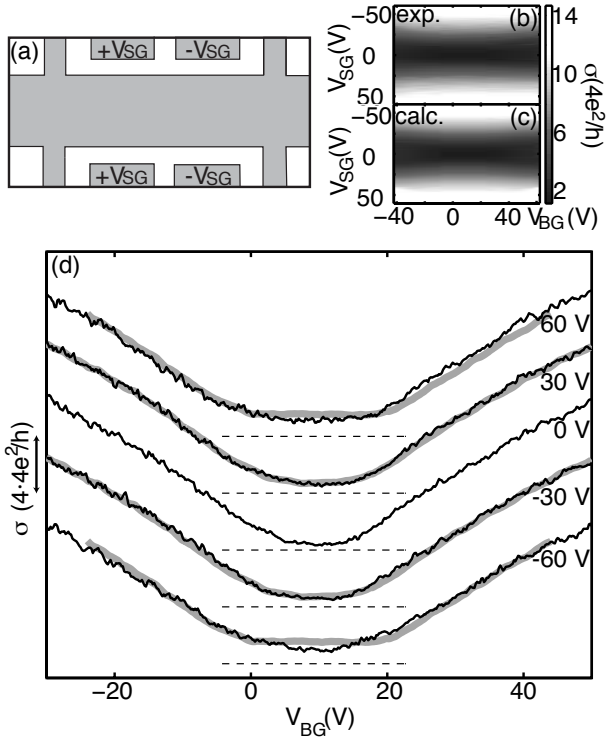}
\caption{\label{fig:PN} (a) Schematic of the Hall bar with indication of the voltages applied to the side gates. An ac current of 100~nA rms is used for these measurements. (b),(c): Conductivity as a function of side gate and back gate voltage, measured (b), and calculated (c). (d) Conductivity as a function of back gate voltage for different voltages applied to the side gates. The black curves represent the measured traces, the gray lines the calculated ones. Individual curves are vertically offset by $4\cdot 4e^{2}/h$ for clarity, and the horizontal dashed lines indicate $4e^{2}/h$.}
\end{figure}

Besides having all side gates at the same potential, many other, more complicated configurations are possible with four side gate electrodes. Here we focus on the situation where the in-plane gates are biased such that a p-n-like configuration along the Hall bar arises. The same voltage is applied to side gates on opposite sides of the Hall bar. For both pairs the applied voltage has the same absolute value, but opposite sign [Fig. \ref{fig:PN}(a)]. Fig. \ref{fig:PN}(d) shows the measured conductivity traces as a function of back gate voltage for different applied side gate voltages (black curves). The different traces are vertically offset by $4\cdot 4e^{2}/h$ for clarity, and the dashed lines indicate a conductivity value of $4e^{2}/h$. Applying side gate voltages as described before leads to an increase of the minimal conductivity value and a broadening of the minimum. However, contrary to the case where the same voltage is applied to all four side gates, the position of the minimum remains unchanged. This indicates that both pairs of opposite side gates have the same lever arm on the Hall bar. Fig. \ref{fig:PN}(b) shows the same measurement for many different side gate voltages. It displays the conductivity as a function of the applied side and back gate voltages, dark colors corresponding to low conductivity values.

The same model as discussed above is applied to this case. Two equally weighted sections, the conductivity of each one being calculated as in the case for the same voltage applied to all side gates, are considered to be connected in series. The transition region between these two sections, where both side gate pairs compensate each other, is neglected: 
\begin{equation}
\sigma_{\mathrm{PN}}(V_{\mathrm{SG}},V_{\mathrm{BG}})=2[\sigma(V_{\mathrm{SG}},V_{\mathrm{BG}})^{-1}+\sigma(-V_{\mathrm{SG}},V_{\mathrm{BG}})^{-1}]^{-1} 
\end{equation}
For the determination of the conductivity in each of the two segments, the same coefficients as determined above are used. Only the curve for $V_{\mathrm{SG}}=0$~V has to be replaced by one measured at the same time as the other traces due to an alteration of the sample with time. Thus, there is no free parameter in these calculations. The gray curves in Fig. \ref{fig:PN}(c) show the resulting traces for some side gate voltages, together with the measured curves (black traces). The calculations reproduce the measured curves remarkably well, except close to the charge neutrality point, where the measured conductivity is lower. We attribute this to having neglected the central region between both side gate pairs, where the effects of both side gate pairs cancel each other. Fig. \ref{fig:PN}(c) shows the calculated conductivity as a function of back and side gate voltages, as compared to the measured data plotted in Fig. \ref{fig:PN}(b). The white corner are in a gate voltage regime which is not accessible from the extrapolation of the data taken at $V_{\mathrm{SG}}=0$.  


\subsection{Quantum Hall regime}

\begin{figure}
\includegraphics[width=0.45\textwidth]{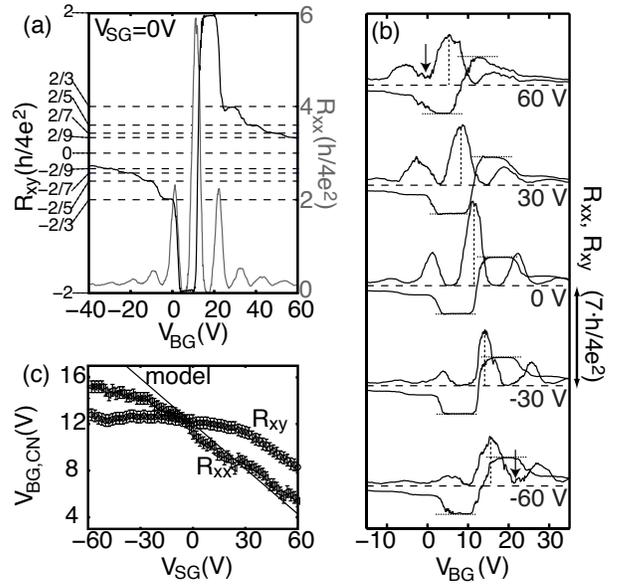}
\caption{\label{fig:8T} (a,b) Magnetoresistance and Hall resistance as a function of back gate voltage, recorded at $B$=8~T and $T$=2~K with an ac current of 100~nA rms. (a) All side gates are at 0~V. The dashed lines indicate the expected positions for plateaus in the Hall resistance for single-layer graphene. (b) Different curves correspond to different voltages applied to all four side gates, and are offset by $7\cdot\frac{h}{4e^{2}}$ for clarity. The vertical dashed lines indicate the center of the middle peak in the longitudinal resistance. The short horizontal dashed lines mark a resistance value of $\pm \frac{h}{2e^{2}}$ where the first plateau in the Hall resistance is expected. (c) Back gate voltage position of the charge neutrality point measured in longitudinal and Hall resistance. The solid line indicates the charge neutrality point shift as calculated by the model described in section \ref{sec:SG} (offset arbitrarily chosen).}
\end{figure}

A direct consequence of the linear dispersion of charge carriers in graphene can be observed in the quantum Hall effect. Unlike in the case of conventional semiconductor interfaces, the Hall conductance plateaus measured for graphene lie at half-integer values of $f\cdot e^{2}/h$, where $f=4$ is the degeneracy factor. \cite{novoselov_nature2005, zhang_nature2005} This can be seen in Fig. \ref{fig:8T}(a), displaying the magnetoresistance (black) and Hall resistance (grey) as a function of applied back gate voltage at fixed magnetic field $B=8$~T and $V_{\mathrm{SG}}=0$~V. The dashed lines represent the expected values for the Hall plateaus, and coincide with the measured ones within the experimental resolution. 

In Fig. \ref{fig:8T}(b) we show the same data upon application of a voltage applied to all four side gates at a fixed magnetic field $B=8$~T. The different curves correspond to different $V_{\mathrm{SG}}$, and are vertically offset for clarity. An applied side gate voltage disturbs the regular shape of longitudinal and Hall resistance curves. Also the back gate position of Hall plateaus and minima in the magnetoresistance shift differently with gate voltage.  The charge neutrality point as extracted from the maximum in the magnetoresistance or the zero of the Hall effect coincide only at $V_{\mathrm{SG}}=0$~V. The magnetoresistance predominately probes the area between the voltage probes. The Hall signal, on the other hand, mainly depends on the density in the Hall cross. The application of a side gate voltage, which has a pronounced effect along the edges of the Hall bar provokes a stronger shift in $V_{\mathrm{BG}}$ of the longitudinal resistance than in the Hall resistance. For large enough side gate voltages, plateaus in $R_{\mathrm{xy}}$ no longer correspond to minima in $R_{\mathrm{xx}}$. If $V_{\mathrm{BG}}$ has the opposite sign than $V_{\mathrm{SG}}$, the longitudinal resistance is lifted compared to the case where the back gate voltage has the same polarity than $V_{\mathrm{SG}}$. In this case, if $V_{\mathrm{SG}}$ is large enough, the resistance minimum no longer reaches zero [arrows in Fig. \ref{fig:8T}(b)]. This could be due to the creation of p-n interfaces, along which counter-circulating edge states run in parallel and can mode mix, therefore leading to a finite resistance. This is analogous to what has been observed by Williams et al. \cite{williams_science2007} for top gate created p-n junctions. Finally, a sufficiently high side gate voltage destroys the plateaus in the Hall resistance. Our range of side gate voltages is not large enough, such that different quantum Hall states were to exist along the Hall bar edge and in the weakly gated region of the Hall cross.

Fig. \ref{fig:8T}(c) compares the back gate voltage position of the charge neutrality point $V_{\mathrm{BG,CN}}$ measured in longitudinal and Hall resistance with the shift predicted by the model described above. The resistance curves are averaged over 1V in back gate voltage, and the charge neutrality point position is then determined by searching the maximum, or the zero point respectively, of the smoothed $R_{\mathrm{xx}}$ and $R_{\mathrm{xy}}$ traces. The solid line indicates the result of the model. The slope of the line is obtained by averaging the relative lever arms of both channels weighted by their widths: $\alpha_{\mathrm{tot}}=\beta\cdot\alpha_{1}+(1-\beta)\cdot\alpha_{2}$. This is in reasonable agreement with the back gate voltage position of the charge neutrality point as extracted from the maximum of the magnetoresistance which is also thought to probe the entire sample. The shift of the charge neutrality point following an analysis of the Hall resistance is rather small, in tune with the expectation that the side gates have little influence on the density in the Hall cross itself. Deviations from this behavior in both quantities for large side gate voltages are probably related to the electric field extending into the region of the voltage probes.


\section{Conclusion}

In summary, we have shown magnetotransport measurements on a single layer graphene Hall bar with additional graphene side gates. The effect of a voltage applied to all side gates in the absence of a magnetic field can be explained by a model taking into account two parallel conductors, being differently influenced by the side gates. The obtained average penetration depth of the side gate created field into the flake is $\approx90$~nm. This makes side gates good candidates for the local tuning of nanostructures such as quantum point contacts and quantum dots. In the quantum Hall regime, an applied side gate voltage affects the longitudinal and Hall resistance differently due to the different overall lever arm in the probed regions.

\begin{acknowledgments}
We acknowledge stimulating discussions with K.S. Novoselov.
Support by the ETH FIRST lab and financial support from the Swiss Science Foundation (Schweizerischer Nationalfonds) are gratefully acknowledged. 
\end{acknowledgments}



\begin{thebibliography}{99}

\bibitem{novoselov_pnas2005}
    K. S. Novoselov, D. Jiang, F. Schedin,T. J. Booth, V. V. Khotkevich, S. V. Morozov, and A. K. Geim, Proc. Natl. Acad. Sci. USA {\bf 102}, 10451 (2005)
    
\bibitem{novoselov_nature2005}
   K. S. Novoselov, A. K. Geim, S. V. Morozov, D. Jiang, M. I. Katsnelson, I. V. Grigorieva, S. V. Dubonos, and A. A. Firsov
, Nature {\bf 438}, 197 (2005)

\bibitem{zhang_nature2005}
   Y. Zhang, Y. Tan, H. L. Stormer, and P. Kim, Nature {\bf 438}, 201 (2005)

\bibitem{lemme_condmat2007}
    M. C. Lemme, T. J. Echtermeyer, M. Baus, and H. Kurz, IEEE Electron Device Lett. {\bf 28}, 283 (2007)

\bibitem{williams_science2007}
    J. R. Williams, L. DiCarlo, and C. M. Marcus, Science {\bf 317}, 638 (2007)

\bibitem{huard_prl2007}
    B. Huard, J. A. Sulpizio, N. Stander, K. Todd, B. Yang, and D. Goldhaber-Gordon, Phys. Rev. Lett. {\bf 98}, 236803 (2007)

\bibitem{trauzettel_naturephys2007}
    B. Trauzettel, D. V. Bulaev, D. Loss, and G. Burkard, Nature Physics {\bf 3}, 192 (2007)

\bibitem{han_prl2007}
    M. Y. Han, B. Ozyilmaz, Y. Zhang, and P. Kim, Phys. Rev. Lett. {\bf 98}, 206805 (2007)

\bibitem{chen_condmat2007}
    Z. Chen, Y. Lin, M. J. Rooks, and P. Avouris, arXiv-condmat: 0701599 (2007)

\bibitem{graf_prb2007}
    D. Graf, F. Molitor, T. Ihn, and K. Ensslin, Phys. Rev. B {\bf 75}, 245429 (2007)
    
\bibitem{luscher_apl1999}
S. L\"uscher, A. Fuhrer, R. Held, T. Heinzel, K. Ensslin, and W. Wegscheider, Appl. Phys. Lett. {\bf 75}, 2452 (1999)

\bibitem{gupta_nanolett2006}
    A. Gupta, G. Chen, P.  Joshi, S. Tadigadapa, and P. C. Eklund, Nano Lett. {\bf 6}, 2667 (2006)

\bibitem{ferrari_prl2006}
 A. C. Ferrari, J. C. Meyer, V. Scardaci, C. Casiraghi, M. Lazzeri, F. Mauri, S. Piscanec, D. Jiang, K. S. Novoselov, S. Roth, and A. K. Geim, Phys. Rev. Lett. {\bf 97}, 187401 (2006)

\bibitem{graf_nanolett2007}
    D. Graf, F. Molitor, K. Ensslin, C. Stampfer, A. Jungen, C. Hierold, and L. Wirtz, Nano Lett. {\bf 7}, 238 (2007)

\bibitem{blake_apl2007}
    P. Blake, E. W. Hill, A. H. Castro Neto, K. S. Novoselov, D. Jiang, R. Yang, T. J. Booth, and A. K. Geim, Appl. Phys. Lett. {\bf 91}, 063124 (2007)

\bibitem{novoselov_science2004}
    K. S. Novoselov, A. K. Geim, S. V. Morozov, D. Jiang, Y. Zhang, S. V. Dubonos, I. V. Grigorieva, and A. A. Firsov, Science {\bf 306}, 666 (2004)

\bibitem{adam_condmat2007}
    S. Adam, E. H. Hwang, V. M. Galitski, and S. Das Sarma, arXiv-condmat: 0705.1540 (2007)


\end{thebibliography}
\end{document}